\documentclass{article}
\usepackage{spconf,amsmath,graphicx}

\title{LEARNABLE FRONT ENDS BASED ON TEMPORAL MODULATION FOR MUSIC TAGGING}

\name{Yinghao Ma$^{1,2}$
, Richard M. Stern$^{2,3}$}
\address{
  $^1$Centre for Digital Music, Queen Mary University of London, UK
  \\
  $^2$School of Music, Carnegie Mellon University, Pittsburgh, PA, USA\\
  $^3$Dept. of Electrical and Computer Engineering, Carnegie Mellon University, Pittsburgh, PA,  USA\\
    }
    
\begin{document}

\maketitle

\begin{abstract}
While end-to-end systems are becoming popular in auditory signal processing including automatic music tagging, models using raw audio as input needs a large amount of data and computational resources without domain knowledge. Inspired by the fact that temporal modulation is regarded as an essential component in auditory perception, we introduce the Temporal Modulation Neural Network (TMNN) that combines Mel-like data-driven front ends and temporal modulation filters with a simple ResNet back end. The structure includes a set of temporal modulation filters to capture long-term patterns in all frequency channels. Experimental results show that the proposed front ends surpass state-of-the-art (SOTA) methods on the MagnaTagATune dataset in automatic music tagging, and they are also helpful for keyword spotting on speech commands. 
Moreover, the model performance for each tag suggests that genre or instrument tags with complex rhythm and mood tags can especially be improved with temporal modulation.
\end{abstract}
\noindent\textbf{Index Terms}: Temporal modulation filters, music tagging, learnable front end

\section{INTRODUCTION}
Deep neural networks (DNNs) have become the standard algorithms for many machine learning tasks including audio representation learning. Although algorithms based on most human-designed audio features have been superseded by neural networks, the (log-)Mel spectrum is still the most commonly-used front end for auditory tasks, including automatic speech recognition (ASR), music information retrieval (MIR), and sound event detection (SED) etc. 

Nevertheless, DNN methods with (log-)Mel spectra as front ends are not necessarily the best choice for music-related tasks. The Mel spectrum is designed for speech applications, and the parameters of Mel filterbanks need not necessarily be optimal for music. Thus, the dimensionality reduction procedure that turns music audio into Mel spectra may lose some valuable information for music tagging, even though it provides good results in speech processing. As a consequence, there may be some advantages for a machine learning system that adapts to the parameters of features on a particular music task. Researchers have attempted to solve this problem by developing end-to-end systems that directly process waveform input in each frame.
While some have produced remarkable results, 
they typically require a large training dataset and substantial computational resources.

Recent research has focused on designing a data-driven front end in an end-to-end system with auditory domain knowledge. 
One type is based on the first-order scattering transform (FST), which uses a series of data-driven band-pass filter banks to turn the raw waveform into a set of time series responses (\emph{e.g.} \cite{
ravanelli2018speaker, 
zeghidour2018learning, vahidi2021modulation}). Another approach uses the short-time Fourier transform (STFT) to represent the data and uses a set of data-driven filter banks to combine multiple frequency bins into a frequency channel at every time step (\emph{e.g.} \cite{won2020data, 
fu2022fastaudio}).
After spectrum calculation, the energy envelopes in all the frequency channels may be downsamplped to reduce information redundancy, and learnable low-pass filters may needed (\emph{e.g.} \cite{zeghidour2018learning, zeghidour2020leaf}). 
Moreover, there will be a compression step to roughly turn the energy value into a scale based on human perception with non-linear activation functions, which can also be data driven (\emph{e.g.} \cite{schluter2018zero, zeghidour2020leaf}).

Some end-to-end models incorporate modulation filters in their front ends, as modulation may emphasize tremolo and vibrato, and thus may be valuable in distinguishing music texture. 
MusicNN \cite{pons2018end} integrates spectral and temporal convolutional kernels representing timbre and rhythm, respectively.
ModNet and SincModNet \cite{vahidi2021modulation}  combine the learned FST spectra of sinc filter banks with a temporal modulation layer. None of them achieve SOTA performance.
The harmonic-NN \cite{won2020data} embeds spectral modulation by Mel-like spectra, a series of learnable triangular filters in the frequency domain, with harmonic information. It shows promising results for multiple audio tasks, but does not embed temporal domain knowledge. 
There are also some modulation models based on human-designed features which may not be optimized individually for each task. For example, Kingsbury \emph{et al.} \cite{kingsbury1998robust} designed a set of fixed parameters as modulation feature extractors. 
Spectro-temporal receptive fields (STRFs) as convolutional filters have been applied to spectra based based on constant-Q transforms (CQT spectra), and subsequently applied to speech-related tasks \cite{vuong2020learnable, vuong2021application, vuong2021modulation}. 
Scattering-related approaches apply temporal and spectral wavelets as front end, which are helpful for audio classification \cite{
anden2019joint, muradeli2022differentiable}.

Given the relationship between auditory perception and temporal modulation phenomena \cite{ding2017temporal}, we present a model with a learnable module that is capable of extracting such modulation patterns in the spectrum. Our contributions are as follows: (i) the design of a learnable front end with a set of temporal modulation filters, (ii) demonstration that our method achieves SOTA performance in music tagging tasks and performs well in keyword detection, (iii) discussion of the number of temporal modulation filters, and (iv) analysis of the impact of modulation processing for each music tag. 

This paper is organized as follows: Section 2 describes our front end. Sction 3 describes the datasets and evaluation metrics. Section 4 discusses the experimental results along with analysis. Finally, Section 5 contains conclusions.

\section{METHODS}
\subsection{Intuition from Auditory Perception}
\begin{figure}[htbp]
\centering
\vspace{-2 mm}
\includegraphics[width=0.47\textwidth]{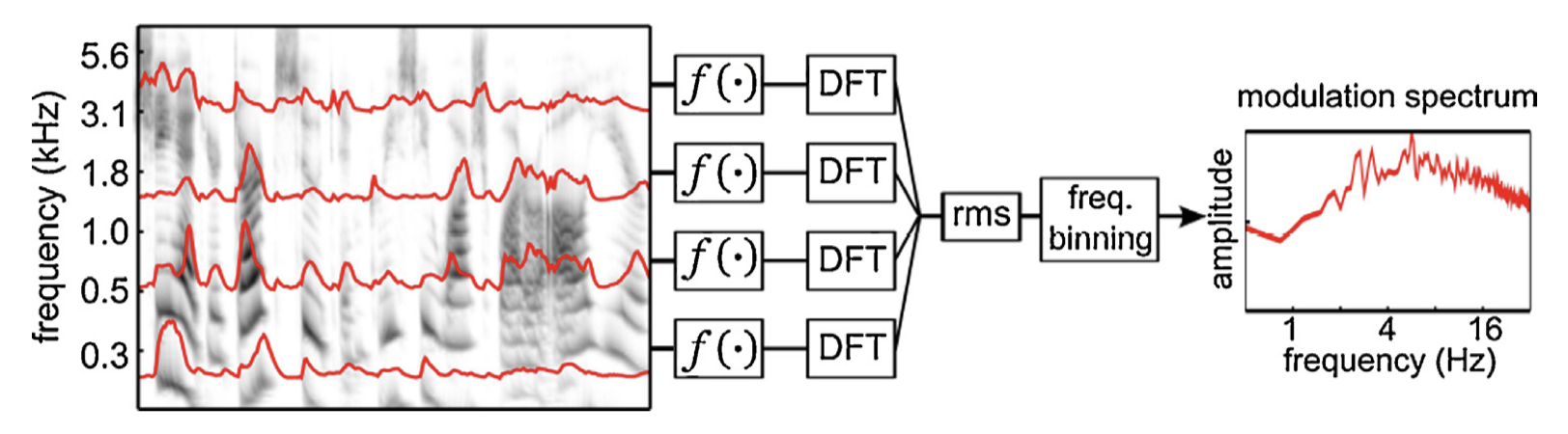}
\vspace{-3 mm}
\caption{Modulation spectrum calculation pipeline. \cite{ding2017temporal}}
\vspace{-1 mm}
\label{fig:perception}
\end{figure}

Figure \ref{fig:perception} shows the procedure that calculates a modulation spectrum. The fundamental spectrum is shown in the gray-scale ﬁgure. Envelopes for selected frequency bands are illustrated by red curves superimposed on it. A nonlinear function $f(\cdot)$ is used for compression. The root-mean-square (RMS) of the Discrete Fourier Transform (DFT) of the envelopes in all channels are the modulation spectrum \cite{yang1992auditory}. The modulation spectrum is believed to roughly reflect how signals are perceived by the human auditory system. It can be viewed as an approximation of the spectrum of neural responses by the auditory nerves or at some higher-level subcortical auditory nuclei, and it can also be roughly viewed as the spectrum of the neural responses summed over the whole neural population of a sub-cortical nucleus. In addition, modulation spectra of various instruments and genres can be quite diverse \cite{ding2017temporal}. Our front-end design is an attempt to mimic such a procedure.



\subsection{Design of the Temporal Modulation Filters}
The front end to be modulated is the learnable Mel-like front end in Harmonic-NN \cite{won2020data}. It is similar to the log-Mel spectrum, but the triangle filters used for spectrum calculation have learnable center frequencies and bandwidths. For the experimental results in Section 4, the first-layer front end applies with the number of harmonic modulations $n$ equal to $6$  provides SOTA results while $n=1$ produces a Mel-like spectrum with learnable center frequency and bandwidth itself without further harmonic modulation.

\begin{figure*}[!tbp]
\centering
\vspace{-6 mm}
\includegraphics[width=0.8\textwidth]{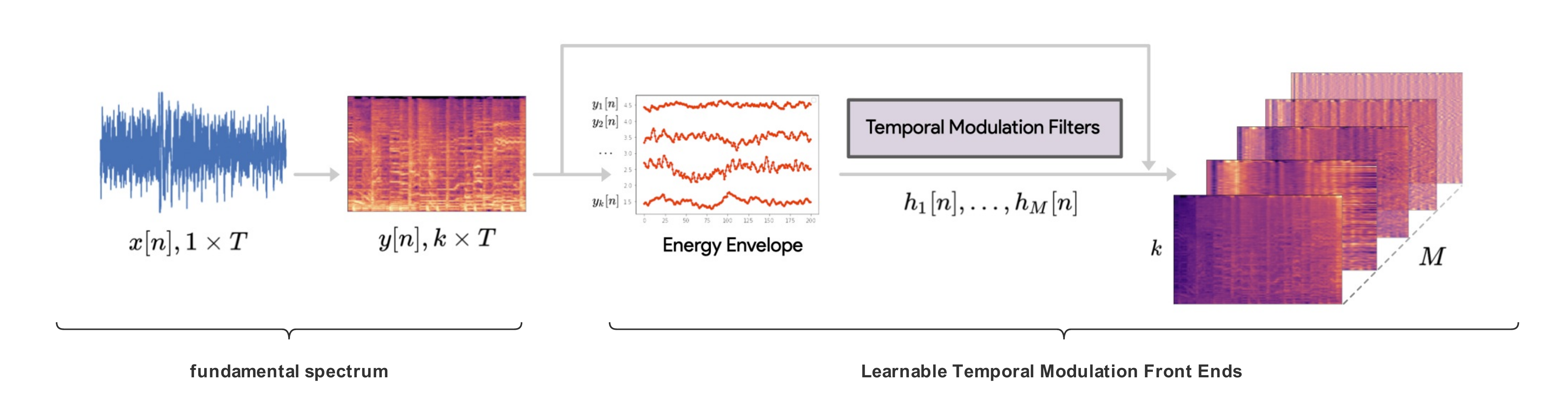}
\vspace{-4 mm}
\caption{Pipeline of the temporal modulation filters}
    \label{fig:TM}
\vspace{-4 mm}
\end{figure*}
Figure \ref{fig:TM} describes the pipeline for the temporal modulation calculations. For each Mel-like spectrum, we calculate the energy envelope of each frequency channel and then apply several temporal modulation filters $h_1[n]$, $h_2[n]$, $\cdots h_m[n]$  for all frequency channels, where $h_i[n]$ is one of the 1-D convolution kernels and a modulation filter truncated by a Hamming window function of size $101$. We constrain the shape of the 1-D CNN kernels to be sinc filters that are bandpass filters in the temporal modulation frequency domain. This is inspired by the observation that some of the learned temporal modulation filters are close to bandpass filters \cite{vaughan1991theory}.
Additionally, the original spectrum will be concatenated with the modulation spectrum. Such a design has two motivations: This is an imitation of the skip connections in ResNet 
to prevent gradients from vanishing. It can also be regarded as adding an all-pass filters to temporal modulation.

We denote the spectrum as $\left[y_1[n]; y_2[n];\cdots y_F[n]\right]$, where $y_f[n]$ is the energy envelope in the $f^{th}$ frequency channel, and where $h_m[n]$ is the $m^{th}$ temporal modulation filters of all $M$ filters. We denote the temporal modulation frequency spectrum as $S$.
Then, the following equation illustrates the modulation operation to evaluate $S$ for all frequency channel $f \in \{1,2,\cdots F\}$ and all modulation filter $m \in \{1,2,\cdots M\}$:
\begin{equation}
\vspace{-2 mm}
    S\left[f, m, n\right] = y_f[n] \ast h_m[n]
    \vspace{-0 mm}
\end{equation}

\subsection{The Structure of TMNN}
Three baseline spectra from earlier time-frequency representations are used: the log-Mel spectrum, the spectral modulation front end from Harmonic-NN  \cite{won2020data} when $n$ is equal to $1$ and $6$. Using these elemental spectra, experiments were carried out with and without the temporal modulation blocks. 
The number of temporal modulations along with the number of harmonics in the harmonic-NN case are regarded as channels for the input of the back end.
A naive back end, consistng of a simple 2-D CNN with 12 layers of ResNet convolution and two fully-connected layers, is applied for all experiments, as the main goal is to evaluate the comparative advantages of our front ends.
We refer to our model, combining the harmonic front end as the first layer, the sinc bandpass temporal modulation filters as the second layer and the naive back end as the rest of the lays, as the Temporal Modulation Neural Network (TMNN).

\section{EXPERMENTAL CONFIGURATION}
\subsection{Dataset}
Automatic music tagging was performed using the MagnaTagATune (MTAT) \cite{law2009evaluation} dataset. This is a multi-label classification that aims to add multiple tags for each given music excerpt. We carry out our experiments on a commonly-used subset of the MTAT dataset as discussed below. 
MTAT consists of 26k audio clips, each of which is 29 seconds on average. 5-second segments (more than two music bars for clips with 4/4 beats) are extracted randomly and independently in each epoch during training. 
The dataset partition is the same as previous studies \cite{
lee2017sample, lee2018samplecnn, 
won2020data}. Only the top $50$ most frequent tags are considered among all the tags. Musical excerpts with none of these 50 tags have been removed from the dataset, which yields about $22$k audio clips. 
These tags include music instruments, vocals, mood, and genre, etc. The evaluation metrics are the area under the curve of the ROC (ROC-AUC) and the area under the precision-recall curve (PR-AUC), which are also the same as in previous studies. 

Keyword spotting was performed using the Speech Commands dataset \cite{warden2018speech}, which consists of about 106k speech excerpts with 35 command classes for limited-vocabulary speech recognition. 
In this dataset, each audio file is about one second long and therefore has at most one label. The dataset sample rate is 16 kHz. The most commonly used version of the dataset, \texttt{torch.utils.data.Dataset}, has been used. All the models are evaluated directly regarding classification accuracy in choosing one of the 35 classes.

\subsection{Training Configuration}
We resampled the music recordings to 16 kHz during the experiments. All front ends, are calculated with 512-point FFTs with a 50\% overlap by default. Therefore, there will be $64$ frames per second of the music clips, and the temporal modulation frequencies range from $0$ Hz to $32$ Hz. For training the models, we used a unified optimization method \cite{won2020data}, a mixture of scheduled ADAM and stochastic gradient descent (SGD). The best model is selected based on the value of the loss function observed in the validation set. The loss function for music tagging tasks is BCELoss, and CrossEntropy for multi-class keyword classification.

\section{EXPERIMENTAL RESULTS}
\subsection{Model Performance on Music Tagging}
\begin{table}[htbp]
  \caption{Music Tagging Results of models with and without temporal modulation on MTAT}
  \label{tab:structure}
  \centering
  \begin{tabular}{l|ll}
    front end types  & ROC-AUC & PR-AUC   \\
    \hline
    Mel spectrum &89.39 &41.17 \\
    Mel spectrum + Temporal &90.62 &43.76 \\
    Harmonic-NN (n=1)  &91.32 \cite{won2020data} &45.99 \cite{won2020data} \\
    Harmonic-NN (n=1) + Temporal &\textbf{91.74} &46.33 \\
    Harmonic-NN (n=6) (SOTA)  &91.41 \cite{won2020data} &46.46 \cite{won2020data} \\
    Harmonic-NN (n=6) + Temporal &91.68 &\textbf{47.13} \\
  \end{tabular}
      \vspace{-5 mm}
\end{table}

Table \ref{tab:structure} shows results using the ROC-AUC and PR-AUC metrics for the three front ends with and without the temporal modulation block described in Section 2. These results suggest that simply adding a temporal modulation block to extract particular patterns in the time domain can improve the performance for all front ends. This implies that not only the harmonic information but also the temporal information of a music clip is informative for music tagging.

\begin{figure}[htbp]
\centering
\includegraphics[width=0.45\textwidth]{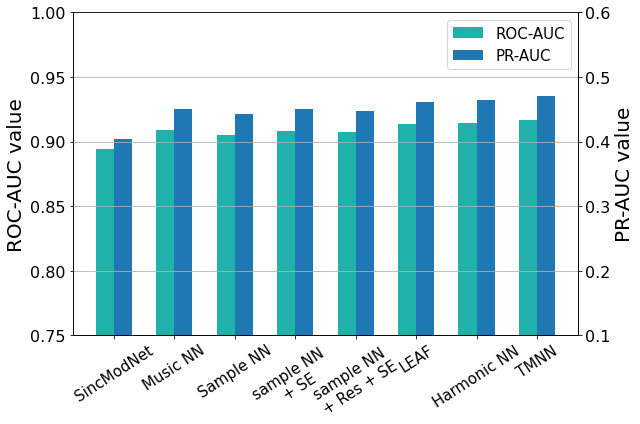}
\vspace{-5 mm}
\caption{Performance of TMNN and other models on MTAT.}
      \vspace{-4 mm}
\label{fig:sota}
\end{figure}

To compare our model to previous studies, we implemented TMNN, harmonic-NN \cite{won2020data}, LEAF \cite{zeghidour2020leaf} and compared their results with other baseline systems including MusicNN \cite{pons2018end}, different versions of simpleNN \cite{lee2017sample, lee2018samplecnn, 
}, along with another temporal modulation SincModNet \cite{vahidi2021modulation}.
Figure \ref{fig:sota} shows results in terms of the ROC-AUC and PR-AUC metrics. The ROC-AUC value of TMNN (n=1) is 0.3\% greater than the current SOTA performance with a p-value of 0.045 and the improvement is not significant for TMNN (n=6). The PR-AUC value of TMNN (n=6) is 0.7\% higher than SOTA with a p-value of 0.036. These imply that the model performance is a little bit better than existing models on both measure. 

\subsection{Consideration of the Number of Modulation Filters}
\begin{figure}[htbp]
    \centering
    \includegraphics[width=0.46\textwidth]{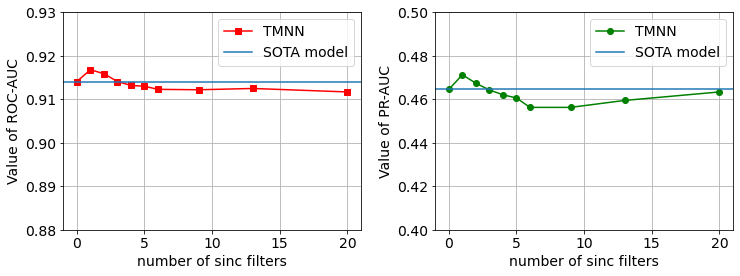}
    \vspace{-4 mm}
    \caption{Performance of TMNN with Different Numbers of Temporal Modulation Filters}
    \label{fig:n_filt}
      \vspace{-6 mm}
\end{figure}
Figure \ref{fig:n_filt} describes performance on the MTAT task as a function of the number of filters used. It shows that performance may decrease slightly when the number of filters in TMNN increases. When the number of filters $n$ is $0$, the model is actually the SOTA model. The best results are obtained when only a single sinc filter is employed. 
For other types of temporal modulation filters like Gaussian or Gabor, the number of filters that provided the best results was always less than or equal to three. 
This might be because the learned focus more on the low modulation frequency which is essential to music classification and lack of diversity when the number of filters is large which might be due to the instability during training.

\subsection{Impact of Temporal Modulation Blocks on All Tags}
When we compare the performance of each music tag with and without modulation blocks, there are some tags that have improvement including the genre tags ``synth,'' ``metal,'' ``Indian,'' instrumental tags ``harp'' and ``flute,'' ``sitar,'' ``harpsichord,'' vocal tags ``male vocal'' and ``female vocal,'' along with music mood tags ``soft'' and ``quiet''.
These results suggest that our model might be helpful for a wide range of music texture tags. 
On the other hand, some tags have worse performance with temporal modulation blocks such as ``beat,'' which is obscure and may not be tagged by some of the players of the music tagging game who define the ground truth. And the tag ``rock'' shows a 1\% decrease in the performance, which might partly be due to the excellent performance the SOTA method already exhibits and the small room for progress. In addition, the tags ``choir,'' ``solo'' and vocals'' decrease significantly might suggest our model can be worse for some type of singing.

\subsection{Model Performance on Speech-Related Tasks}
\begin{figure}[htbp]
    \centering
    \includegraphics[width=0.44\textwidth]{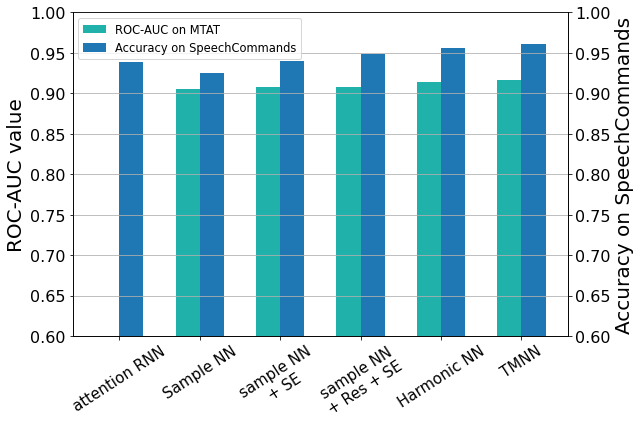}
    \vspace{-4 mm}
    \caption{Keyword spotting accuracy on Speech Commands}
    \label{fig:speech}
    \vspace{-5 mm}
\end{figure}
To determine the extent to which TMNN can be helpful for other auditory signals besides music, we tested the TMNN model on the simple keyword spotting task. The experimental setting for speech is the same as music tagging experiments except the number of output classes in the last layer. The multi-class keywords classification accuracy of TMNN and baselines are shown in Figure \ref{fig:speech}. We can see that TMNN, which emphasizes the temporal modulation effect, can contribute to speech-related tasks as well. Although the input length of the speech commands dataset is only 1 second or so, the temporal modulation may help the back end to process information with more than 0.24 seconds of input and therefore contribute to multi-classification with long-term effects. Such results do not surpass the SOTA approach, as the back end of the SOTA model is typically much more complicated, and the TMNN only includes a naive back end.

\section{CONCLUSIONS}
We introduced learnable front ends that interpret inherent temporal modulation on the spectrum for audio classification based on Mel-like front ends. Inspired by the temporal modulation phenomenon in auditory perception, the proposed structure includes a set of temporal modulation filters that compels the network to capture long-term patterns in each frequency band while preserving spectral-temporal locality. Experimental results with a simple back-end CNN show that our front ends surpass the state-of-the-art baseline method on the MagnaTagATune dataset in automatic music tagging and are also helpful for keyword spotting on speech commands. We also considered the temporal modulation frequency response to explain some failures of the TMNN model. The results suggest that the best modulation filters are approximately low-pass when the number of temporal modulation filters is small. Finally, we evaluated the performance of the TMNN model for each of the tags. The results imply that genre tags with complex rhythms like ``Indian'' and some mood tags like ``soft'' and ``weird'' show some improvement. These results suggest that temporal modulation can provide gains in performance for the music tagging task with some long-term effects compared with the SOTA model.

\section{ACKNOWLEDGEMENTS}
Yinghao Ma was supported by a Carnegie Mellon University GSA/Provost GuSH Grant for this project. He is currently a research student at the UKRI Centre for Doctoral Training in Artificial Intelligence and Music, supported by UK Research and Innovation [grant number EP/S022694/1]. We would like to thank Dr. Roger Dannenberg, Dr. Bhiksha Raj, Dr. Emmanouil Benetos, and Mr. Riccardo Schulz for providing guidance or feedback throughout this work.

\bibliographystyle{IEEEbib}
\bibliography{strings,refs}

\end{document}